# Ten Questions in Lifelog Mining and Information Recall

An-Zi Yen[1], Hen-Hsen Huang[2,3], Hsin-Hsi Chen[1,3]
[1]Department of Computer Science and Information Engineering,
National Taiwan University, Taipei, Taiwan
[2]Department of Computer Science, National Chengchi University, Taipei, Taiwan
[3]MOST Joint Research Center for AI Technology and All Vista Healthcare, Taiwan
azyen@nlg.csie.ntu.edu.tw, hhhuang@nccu.edu.tw, hhchen@ntu.edu.tw

## Abstract

With the advance of science and technology, people are used to record their daily life events via writing blogs, uploading social media posts, taking photos, or filming videos. Such rich repository personal information is useful for supporting human living assistance. The main challenge is how to store and manage personal knowledge from various sources. In this position paper, we propose a research agenda on mining personal knowledge from various sources of lifelogs, personal knowledge base construction, and information recall for assisting people to recall their experiences.

## 1 Introduction

The concept of lifelogging was first introduced in the proposal of Memex [Bush, 1945], a hypothetical system allowing a person to store all the knowledge collected in her/his lifetime and to organize the records in a hypertext form. Bush regarded the Memex system as an "*enlarged intimate supplement to one's memory*", which can be consulted speedily and flexibly. In this sense, lifelogging is a process to actively capture and record daily experiences of an individual, namely, a lifelogger. In general, the collected life experiences called lifelogs are typically stored in digital formats nowadays, and can be viewed as personal big data that keep personal life events.

With the advance in digital technology, lifelogging becomes more feasible. The possible sources of lifelogs include wearable cameras, wearable sensors, smartphones, and computers. For example, lifelogs could be messages to communicate with other people, location cues captured by GPS, physiological state via biometric sensors, digital photo albums, blog posts with photos, and video clips captured by an action camera. These personalized multimedia data present various aspects of an individual's daily experiences, offering rich repository information for a variety of personal lifelogging applications, including lifestyle understanding [Doherty *et al*., 2011], diet monitoring [Maekawa *et al*., 2013], human memory recall assistance [Woodberry *et al*., 2015], and visual lifelog retrieval [Chu *et al*., 2019; Fu *et al*., 2019].

Typically, data in a timeline can be segmented into a series of personal life events of various durations. With such a huge amount of life event archives, it is not easy to access the desired information without an efficient system for analyzing, indexing, and organizing the collected data. Besides, the lack of contextual information as well as the noise in the records can be a challenge to construct such a system. To manage these vast personal data efficiently, lifelogs should be stored as structured data for easy access.

Knowledge base has been used to organize, manage, and retrieve structured information. Large-scale knowledge bases such as Freebase and DBpedia contain real-world facts, and provide powerful structural information to various natural language processing applications, e.g., question answering and relation extraction. In order to store and organize lifelogs, constructing a personal knowledge base for an individual is necessary. Generally, the facts in a knowledge base are about places, celebrities, or public events. In other words, the entities, which are not "famous enough" or do not belong to world knowledge, are excluded from the knowledge base. In contrast, the personal knowledge base is used to store the daily life experiences and the facts related to an individual. In this paper, we define the concept of personal knowledge, and discuss how to construct a personal knowledge base from a variety of sources. Personal lifelogging applications can be formulated as: a person uses the lifelogging system to record personal information for her/his own benefits. Constructing personal knowledge bases for individuals is important for the applications of information recall and living assistance. Information recall forms another challenging task, which aims to find the correct facts at the correct timing. We will explore the scenario of the use of information recall service, as well as possible challenges. A total of 10 research questions will be raised and discussed in detail.

The contributions of this position paper are threefold. (1) We point out the concept of personal knowledge from a variety of sources. (2) We comprehensively identify the challenging issues involved in the construction of a personal knowledge base and its connection to a world knowledge base. (3) We explore the application of information recall based on a personal knowledge base.

The rest of this paper is organized as follows. In Section 2, we discuss the potential sources of lifelogs. Section 3 presents the personal knowledge base construction according to different granularities of lifelogs from heterogeneous sources.

Section 4 introduces the information recall service and highlights the use of the personal and the world knowledge bases. Section 5 addresses the privacy and ethics issues in lifelogging. Finally, Section 6 concludes the remarks.

## 2 Personal Events in Heterogeneous Lifelogs

Lifelogs can be viewed as personal big data that record personal daily experiences, including the place where an individual visited in daily life, someone who was at a party s/he attended, and the moment when s/he got the job promotion. We define these experiences as personal life events. Table 1 summarizes previous work of personal life event detection, which is a fundamental task of lifelog mining. Most of previous works focus on major life events, which seldom occur in daily life, such as having a child, marriage, and graduation. Although these works achieve good performance on detecting personal life event, these personal life events are not enough to support the personal knowledge base construction. In this section, we discuss different sources of data that can be employed for lifelogging and introduce the public datasets used for the research of lifelogging.

|  | # of Events | Precision | Recall |
|---|---|---|---|
| Li et al., [2014] | 42 | 62% | 48% |
| Choudhury and Alani [2014] | 5 | 55%~80% | 87%~95% |
| Choudhury and Alani [2015] | 11 | 90% | 69% |
| Dickinson et al. [2016] | 11 | 90.7%~93.5% | 90.6%~93.0% |

Table 1: Previous researches on detecting personal life events.

The traditional way people used to record their life was to write a diary. A diary is composed of a sequence of sentences that record what happens in a period of time. They usually include the author's personal thoughts and feelings. In recent years, recording personal daily life via wearable devices and social media platforms such as Twitter, Facebook, and Instagram becomes very popular. The wearable sensors provide locations and biometrics information of human in daily events. The wearable cameras such as the GoPro action camera create a new media type, namely Video Weblog (VLog). Compared with wearable devices, social media platforms provide large scale text-based lifelogs and images.

Based on the lifelog categories defined by Machajdik et al. [2011] and Gurrin et al. [2019], we classify the source of lifelogs into six parts, including activity visual capture, multimodal data creation, communication activity, personal biometrics, location information, and computer activity. We also define the meaningful unit for each source.

**Activity visual capture:** Currently, wearable cameras for lifelogging, such as SenseCam, Video glass, and Go-Pro, can consequently and actively capture images and videos of what users see in front of them at a frequency of several times per minute. Intuitively, a meaningful unit of a video is a video clip that records the daily life of an individual. In a video clip, the visual and audio data imply the activities of lifelogger. The visual data represent the actions of lifelogger. For example, the object concept "food" might give us a clue for knowing the image is about the semantic activity "eating" of the lifelogger. A steering wheel in the video implies that the lifelogger is driving a car. On the other hand, people usually talk about where they are, what they see, what they are doing, and feelings when they film their Vlog. Both visual and audio data are important sources to extract personal knowledge.

**Multimodal data creation:** Apart from lifelogging via wearable cameras, people are used to write blogs or log their life in social media such as Twitter, Facebook, and Instagram. People create the multimodal data include a combination of text and images to describe their life experiences. A post is a meaningful unit of activity on social network, e.g., a tweet or a blog. We can extract the information of "Who did What to Whom Where When and How" from textual data to compose the personal knowledge of lifelogger. The accompanying image in textual data provide auxiliary information for recognizing activities.

**Communication activity:** With the rapid development of mobile technology, people often use mobile phones to communicate with others. The communication activity represents our electronic communications, such as phone calls, text messages, and emails which can be formed part of a lifelog. The meaningful unit of communication activity is a text we send and receive.

**Personal biometrics:** Wearable sensors record bio-signal for monitoring human everyday performance, e.g., heart rate, calorie burn, steps, and sleep duration. Comparing with the lifelogs we mention above, biometrics information is continuously and actively recorded for a long time. The meaningful unit of biometric information is physical data recorded by wearable sensor at each time interval.

**Location information**: The location information logs the motion of lifelogger, such as GPS data, acceleration, and movement are continuously and passively captured by smartphone with built-in GPS, accelerometers, compass, and camera. The meaningful unit of location information is the coordinates of a location captured by wearable sensor.

**Computer activity:** In addition to communications, the activities on our laptop and desktop computers, which can be monitored such as keystrokes, documents saved, web pages browsed, and YouTube videos watched. The meaningful unit of computer activity log is every step of our activity on the computer. The research of lifelogging relies on public datasets for studying in public for a long time. In order to obtain sufficient information for identifying daily living of an individual, lifelog data should be captured continuously for a long enough time span. Therefore, the data collection would be time-consuming. We list the datasets that contain daylong lifelog records in Table 2.

Most datasets captured by wearable devices contain images and videos. In addition to visual information, Li and Cardie [2014] crawl dataset from Twitter. They collect English tweets and detect important tweets to construct personal timeline. Multimodal lifelogging attracts attention in recent studies. The last row of Table 2 shows a multimodal lifelog dataset, which consists of both textual and visual information shared on the social media platform.

| Dataset | Lifelog Categories | Subjects | Size | Period | Description |
|---|---|---|---|---|---|
| All I Have Seen (Jojic et al., 2010) | Activity visual capture | 1 | 45,612 images | 19 days | The dataset contains personal daily activities, which took place around work office, playgrounds, supermarkets, campus, etc. |
| First-Person Social Interactions Dataset (Fathi et al., 2012) | Activity visual capture | 8 | Day-long videos | 42 hours | The videos are captured by head-mounted GoPro camera at Disney World, with actions performed annotated |
| Twitter Timeline Generation Dataset (Li and Cardie et al., 2014) | Multimodal data creation | 20 ordinary users and 20 celebrities | 36,520 English tweets from the ordinary users. 132,423 English tweets from the celebrities. | 637 days | They focus on detecting important event to construct a person's life history from Twitter stream. |
| Egocentric Dataset of the University of Barcelona (Bolanos et al., 2016) | Activity visual capture | 4 | 4,912 images | 2 days | The dataset contains daily activities captured by the wearable camera Narrative, such as shopping, eating, working, with object labels and segmentations annotated |
| R3 Dataset (Molino et al., 2018) | Activity visual capture | 57 | 1.5M images | 1,723 days | Only extracted visual features are released because of privacy issues. |
| The NTCIR Lifelog Dataset Series (Gurrin et al. 2016; 2017; 2019) The ImageCLEF LifeLog dataset Series (Dang-Nguyen et al., 2018; 2019) | Activity visual capture, Personal biometrics, and Location information | 2~3 | 80K images | about 40 days | Each dataset contains various sensor outputs such as semantic content from mobile devices, heart rate, calorie burn, and steps. |
| Lifelog Dataset (Yen et al., 2019a; 2019b) | Multimodal data creation | 18 | 25,344 Chinese tweets | 2009~2017 | The dataset is labeled with the subject, predicate, object, and time of each life event. The explicitness of the life event (i.e., explicit or implicit) is also labeled. |

Table 2: Existing lifelog datasets.

In summary, lifelogs can be collected from various sources. The first research question arises (**RQ1**): What kind of personal information has to be logged that can support lifelogging application?

## 3 Personal Knowledge Base

The issues of constructing a personal knowledge base include (1) which lifelog should be recorded, (2) how to construct a personal knowledge base from different sources of lifelogs, and (3) how to connect personal knowledge to the world knowledge. In Section 3.1, we discuss the relation between a personal knowledge base and a world knowledge base. The issue of connecting them is also discussed. In Section 3.2, we explore the challenges of constructing personal knowledge bases from various sources of lifelogs.

### 3.1 From World Knowledge Base to Personal Knowledge Base

A personal knowledge base consists a set of personal knowledge represented in a structured format. In general, the facts stored in a world knowledge base are usually represented in the form of the triple (subject, predicate, object), where subject and object are entities and predicate is a relation between entities. For example, (Barack Obama, place_of_birth, Honolulu) means that the birth place of Barack Obama is Honolulu. However, some facts are time-variant, such as presidents of countries or CEOs of companies. Those involve understanding temporal relations. Therefore, the triples in temporal knowledge base are augmented with time information, e.g., (Barack Obama, PresidentOf, US, [2009~2017]). The representation of facts in personal knowledge base might be similar to that of temporal knowledge base because a personal life event also requires time information to indicate when the event happens.

No widely-acknowledged taxonomy of relations in personal knowledge bases is defined yet. Thus, the second research question arises (**RQ2**): How to conduct a universal taxonomy of the relations that covers most personal knowledge?

The temporal information of personal knowledge can be further categorized into one-time, periodical, and durational. For example, (User, Motion, McDonald's, 2018-12-23) is a one-time event denoting User went to McDonald's on December 23, 2018. (User, Travel, Japan, [2017-01-15~2017-01-25]) represents that User travels around Japan for 10 days, which is a durational event. Besides, the variety of time expression, e.g., explicit expression "April 1, 2019" and implicit expression "Rio Summer Olympics opening ceremony", makes capturing events and their time more challenging.

Furthermore, the personal life event can be divided into two categories, static life event and dynamic life event. Static life events represent the "facts" around our life, which are similar to the information highlighted in the infobox on the Wikipedia, such as birth, marriage, and graduation. Dynamic life events mean the events that are frequently and repetitively happen in our daily life, like eating, watching TV, visiting a local place, having a talk with friends, and so on. Balog and Kenter [2019] define the concept of personal knowledge base as "*a resource of structured information about entities personally related to its user*", which covers the static life events. The dynamic life events extraction from social media posts to construct personal knowledge base [Yen *et al*., 2019a; 2019b] is also proposed. Besides, some personal knowledge is involved in world knowledge. Therefore, personal knowledge base that contains static and dynamic life events may connect with external knowledge bases.

Figure 1 shows an example of a personal knowledge base connecting external knowledge bases. Each node denotes an entity, and line denotes the relation between two entities. Green line and green node denote static life event, and blue line and blue node denote dynamic life event. The dotted lines are the connection of entities between personal and external knowledge bases. For instance, the hometown where the user was born is a static life event, and the hometown is connected to other entities in the world knowledge base. Specifically, (User, place_of_birth, Tokyo) represents the user was born in Tokyo. The location entity, Tokyo, can be connected to the facts in the world knowledge base, e.g., (Tokyo, country, Japan). On the other hand, "the user bid a bike" is a dynamic life event which connects to the domain-specific knowledge base and e-commerce catalog.

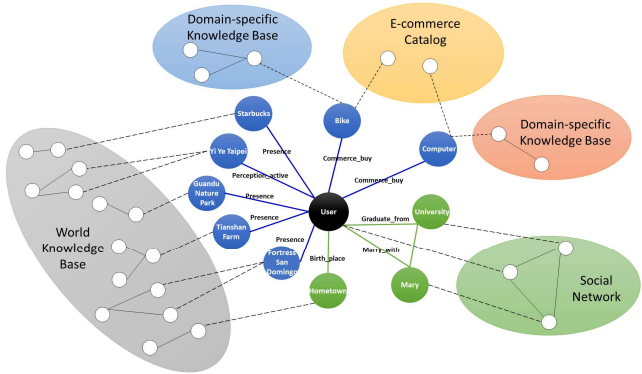

Figure 1: A snippet of personal and external knowledge base.

One of the challenge of constructing a personal knowledge base is linking entity mentions to entries in a given database or dictionary of entities. Traditional entity linking approaches usually utilize resources like alias table or frequency statistics. However, the entities in the user generated texts might not occur in any documents and any Wikipedia pages because of the informal-written problem or the entities not belonging to world knowledge, e.g., the name of the user's pet. The challenge leads to the next research question (**RQ3**): How is a world knowledge base integrated with a personal knowledge base? A critical issue is that how entity linking methods can be utilized on connecting a personal knowledge base and a world knowledge base.

### 3.2 Personal Knowledge Base Construction

Personal knowledge base construction can be considered as a special case of temporal knowledge base construction. Both of them store the facts in the quadruple form (subject, predicate, object, time). Temporal knowledge base construction focuses on extracting subject, predicate, object, and time from textual data. By contrast, constructing a personal knowledge base extracts personal live events from various sources of lifelogs, including textual data, visual data, metadata, and location information.

**Extracting Personal Knowledge from Textual Data**
Extracting personal knowledge from textual data is a task of information extraction. Traditionally, information extraction focuses on detecting and classifying mentions of people, things, locations, events, and other pre-specified types of concepts. However, personal knowledge extraction requires filtering the text that mentions a world event or an opinion of public issue at first, which is irrelevant to daily life experience of an individual. In addition, life events in a text description might be expressed implicitly [Yen *et al*., 2019a; 2019b], recognizing implicit life events and representing them explicitly is a challenge issue.

**Extracting Personal Knowledge from Visual Data**
Recently, the advance in wearable technology has made life-logging more popular. In addition to detecting and extracting personal knowledge on social media posts, video is also an important source of lifelog data. Vlog, a form of blog via recording video, which contains a lot of audio and visual information to reveal the life events of a user, can be used to construct personal knowledge base. As we mention before, the meaningful unit of video is a video clip, the consideration leads to the fourth research question (**RQ4**): How to identify video scene boundary before extracting personal knowledge from the video?

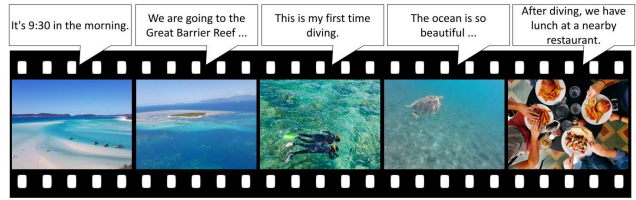

Figure 2: A toy example of a Vlog.

Figure 2 shows a toy example of vlog that the user went to the Great Barrier Reef to dive. They saw a sea turtle, and then went for lunch. Precisely extracting personal knowledge from videos requires both of visual and textual information. Taking the first two scenes as examples, visual information is insufficient for covering the personal knowledge, while the subtitles consist of the time and location information. And the last two scenes, the personal knowledge about what the user saw in the ocean and what the user ate after diving can be extracted by combining both visual and textual information.

For constructing the personal knowledge base via visual data, we formulate two possible tasks, namely visual lifelog object detection and visual lifelog activity recognition. The former task focuses on detecting objects in the image (or video). The latter task, which is regarded as a special case of human activity recognition with the first-person point of view, is aimed at automatic recognition of lifelog data in terms of activities of daily living, e.g., eating, shopping, cooking, relaxing, etc. Thanks to the advance in computer vision (CV) technologies, we could identify the place and recognize multiple objects in an image with CV models trained for place recognition and common objects recognition, respectively. The visual concepts extracted from CV model could provide a shallow semantic interpretation for each image. For instance, in the picture on the right, we can recognize the burger and fries on the plate through the visual lifelog object recog-nition model, and then use the visual lifelog activity recognition model to recognize the user is eating. Given the activity and object recognized by CV models, the personal knowledge can be represented in the quadruples (User, Ingestion, burger, Date) and (User, Ingestion, fries, Date). However, the CV models might generate false detection, automatically filtering out the irrelevant visual concepts is an important challenge for constructing the personal knowledge base from visual data.

The main challenge of visual lifelog activity recognition is the semantic gap between the visual concept and the textual description of events for accessing multimedia lifelog. However, most multimodal CV models are trained on modeling rather low-level descriptions such as concrete objects and places [Zhou *et al.*, 2017]. This problem leads to the next research question (**RQ5**): How to associate extracted semantic contents with a more high-level description of an image? For example, it might indicate that the user is working when the video scene is a desk with a laptop computer on it.

**Extracting Personal Knowledge from Numerical Data**
Numerical data such as the personal biometrics and location information provides other lifelogs. Location information indicates where the user has been. We can find the name of the place through the GPS coordinate, and transform it into knowledge base facts. However, the granularity of GPS information may be too fine or too rough to record. For instance, when shopping on the road, we may obtain a lot of GPS information via wearable sensor, but we only need to record the user have been to a certain road or some stores. Similar to location information, we also need to extract important personal knowledge from computer activity logs, like what website the user visited, rather than what button the user clicked.

On the other hand, the biometrics information such as heart rate, calorie burn, and steps is a continuously recorded digital data. Biometrics information represents our physical state today and is also worth incorporating as a part of the presentation of personal knowledge, e.g., the personal knowledge (User, Motion, School, 2019-06-03, heart rate = 80 bpm) indicates the heart rate was 80 bpm when the user went to school on June 3, 2019. The research question has yet to explore (**RQ6**): How to log important location information, computer activity, and biometrics information at an appropriate level for representing personal knowledge?

To sum up, personal knowledge can be extracted from textual, visual, and numerical data. The seventh research question arises (RQ7): How to transform the personal knowledge extracted from different sources to a uniform structured representation?

**Extracting Personal Knowledge with Emotions**
Previous lifelogging research only focuses on capturing the "facts" around the lifelogger's life. Thoughts and feelings identification are entirely missing in current lifelogging systems. Identifying emotion of the corresponding life event enables the lifelogging system to trace the lifelogger's emotional changes in a given life event and facilitates further applications, such as recommendation. In this case, the personal knowledge is incorporated with emotion as timestamped subject-relation-object-time-emotion facts. The implementation of personal knowledge emotion identification brings the following research question (**RQ8**): How can a lifelogging system identify the lifelogger's emotion in a personal knowledge? This question is related to the aspect-based sentiment analysis, which is aimed at identifying the specific sentiments towards different aspects of an entity, such as food and service in restaurant reviews. In previous aspect-based sentiment analysis systems, the aspects are a small set. By contrast, the aspects of sentiment in personal knowledge are much more diverse.

## 4 Information Recall

People often forget something over time, such as forgetting the names of exact entities in their life events and encountering situations that require recalling the experiences in their daily life. Information recall support for people at the right time and at the right place is emerging. Lifelogs could help us recall a specific piece of past experiences in detail, e.g., recollecting the name of the place where we visited or who was at a party we attended. In addition, lifelogs could also relive past experiences for emotional or sentimental reasons, and are expected to retrieve the specific digital item or information such as documents, emails or websites.

As we mention in Section 3.2, the lifeloggers might log their life by images or videos. Information recall on the visual data was formulated as the task of image/video retrieval or visual question answering (VQA) in previous work [Chu et al., 2019; Dang-Nguyen et al., 2018; 2019; Fu et al., 2019; Gurrin et al. 2016; 2017; 2019]. However, the current studies of VQA are still limited to query the object properties such as an object's color, shape, and region in visual data. For information recall, the question is more likely to be about abstract concepts, e.g., dining, shopping, and driving and so on.

Jiang *et al*. [2017] construct the MemexQA dataset that contains questions about real-world personal photo albums, and propose a multimodal end-to-end neural network model for memory recall only by contextual understanding of personalized data. The information recall service considering multimedia information and world knowledge has yet to investigate. Compared to images, videos provide more information. For example, the query "*when did we last go to the Great Barrier Reef*" requires to reason on video clip, and even world knowledge. On the other hand, people usually describe and query their questions with textual expressions, while the lifelogs via wearable camera are visual data. That is, the challenging issue of image recall reducing the semantic gap between visual and textual domains for effectively querying lifelogs constructed from images and videos.

Extracting semantic contents from visual data and constructing personal knowledge base can support information recall on multimedia lifelogs. An information recall system may reactively accept service requests or proactively provide services. Figure 3 shows a scenario of the use of information recall service. Figures 3(a) and (b) show two service designs in the reactive mode and the proactive mode, respectively.

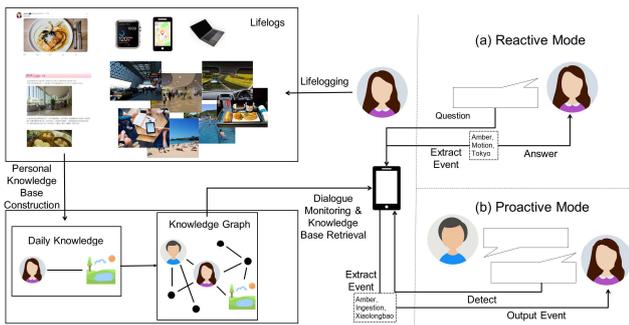

Figure 3: Scenarios of Information Recall Support.

In the reactive mode, users directly ask the system questions in order to recall the memory, which can be regarded as an application of knowledge based question answering (KBQA). Storing life events in a knowledge base benefits from the progress of previous research such as complex question answering over knowledge base [Luo *et al.*, 2018], which have been explored in recent years. In contrast to general KBQA task, the information recall service allows a user to query experiences in natural language over the personal knowledge base. For example, the user might ask "Which monument in Paris have I visited?" or "Which monument have I visited was established around 1600?" Directly retrieving the blogs or social media posts cannot provide the correct answer to the questions containing personal knowledge with world knowledge. This example shows that personal knowledge base construction is necessary for information recall. In addition, to answer these questions requires more than one factual triple and world knowledge. Therefore, the issue of connecting personal knowledge base and world knowledge base is important for supporting information recall. It is worth noting that the user may also query the personal knowledge related to duration or frequency. For example, the user may want to query how many days she spent on traveling around the island by riding a bicycle last year. For a user traveling to Japan many times, she may want to know which hotel she stayed during her second time in Tokyo.

Given above issues, we raise the next research question (**RQ9**): Do the existing methods of question answering systems support information recall service?

The application scenario in Figures 3(b) belongs to the proactive mode. The user talks to her friend to share her life experiences. However, she could not remember the name of the exact entities. The user's situation automatically triggers the information recall service to retrieve the event by searching the knowledge base of her experience using the information in their conversation. This goal raises a new research question (**RQ10**): What is the right time that the system should be triggered in the conversation and provide information recall services? One possible indicator of people requiring information recall is hesitation in speech. When people encounter difficulties in information recall while speaking, hesitation is a common phenomenon. Wang et al. [2018] propose a model for detecting hesitation in conversations by using discourse markers.

## 5 Solutions to Ethical and Privacy Issues

With advances in sensors for sensing the person as well as environment, and cheaper computer storage, technologies enable us to capture lifelogs easily. However, the lifelog might contain a complete digital trace of personal life. It poses new concerns about the societal acceptance, privacy, and the data ownership [Gurrin *et al.*, 2014a]. Ethical and privacy issues attract much attention in lifelogging, resulting in restricted applications. Solutions to ethical issues are mandatory for the progress of lifelogging. The lifelogs gathered from a variety of sources contain personal information that the lifeloggers might not be willing to share. Gurrin *et al.* [2014b] discuss a privacy-aware lifelogging framework with a privacy-by-design approach. Zhou and Gurrin [2012] ask participants about their thoughts of lifelogging. They feel uncomfortable wearing devices. More importantly, the major concern of people is that their personal data may be disclosed to the public. There are also challenges about who owns the data and who can access the data. Storing sensitive data in a cloud-based service would not be acceptable.

Different from the previous concerns, the public social media posts and blog posts are utilized to explore models for personal knowledge base construction [Yen et al., 2019a; 2019b] and image recall [Chu et al., 2019], respectively. The innovative strategies are conducted on lifelogging applications with the data that users are willing to share. That can avoid offending privacy issues.

## 6 Conclusion

This paper summarizes the concepts of the personal knowledge base and proposes a research agenda of personal knowledge base construction and the application on information recall. First of all, we investigate the possible sources of lifelogs data and discuss the meaningful unit of each lifelogs. The relation between personal knowledge base and world knowledge base is also presented. Then, we explore the research topics and challenges of constructing a personal knowledge base from each source. Finally, since the main purpose of a lifelogging is to trace a person's activities to provide living assistance for an individual, we investigate the scenario of the use of information recall service. The information recall system can reactively or proactively assist users to recall their experiences.

In conclusion, many research topics in personal knowledge base construction are worth investigating, e.g., the personal knowledge representation, the methods of personal knowledge extraction from multimedia lifelogs, and entity linking. Considering privacy issues of lifelogging, personal information encryption is also a major topic in the future.

## Acknowledgments

This research was partially supported by Ministry of Science and Technology, Taiwan, under grants MOST-106-2923-E-002-012-MY3, MOST-109-2634-F-002-040-, MOST-109-2634-F-002-034-, MOST-108-2218-E-009-051-, and by Academia Sinica, Taiwan, under grant AS-TP-107-M05.